\documentstyle[sprocl]{article}

\def\be{\begin{equation}}
\def\ee{\end{equation}}
\def\bea{\begin{eqnarray}}
\def\eea{\end{eqnarray}}

\begin{document}

\begin{flushright}
{SNUTP/97-066\\hep-th/9708038}
\end{flushright}

\vspace{5mm}

\title{Thirring Model in Lower Dimensions: Nonperturbative 
Approaches\footnote{To appear in the proceedings of the International
Workshop on Prespectives of Strong Coupling Gauge Theories(SCGT96),
Nagoya, Japani, November 1996.}}

\author{Yoonbai Kim}

\address{Department of Physics, Sung Kyun Kwan University, Suwon 440-746,
Korea\\ yoonbai$@$cosmos.skku.ac.kr}


\maketitle\abstracts{A reformulation of the Thirring model as a gauge theory
on both continuum spacetime and discretized lattice is reviewed. In (1+1) 
dimensions, our result reproduces consistently the bosonization of the massless
Thirring model. In (2+1) dimensions, the analysis by use of Schwinger-Dyson 
equation is shown to exhibit dynamical fermion mass generation when the number
$N$ of four-component fermions is less than the critical value $N_{\rm cr}=
128/3\pi^{2}$.}

\section{Gauge Theory Formulation of Thirring Model}
\indent\indent  The Lagrange density of Thirring model is given by
\begin{equation}\label{thi}
{\cal L}_{\rm Thi}=\sum_{a}\bar{\psi}_{a}i\gamma^{\mu}\partial_{\mu}\psi_{a}
-\frac{G}{2N}\sum_{a,b}\bar{\psi}_{a}\gamma^{\mu}\psi_{a}\,\bar{\psi}_{b}
\gamma_{\mu}\psi_{b}.
\end{equation}
Here $\psi_{a}$ is treated as a formal one in arbitrary dimensions, but a
four-component Dirac spinor in (2+1) dimensions and $a$, $b$ are summed over
from 1 to $N$.
(For the notations and the detailed calculations, see Ref.[1,2].)
Since the form of four-fermion interaction
term is a contact term between vector currents,\footnote{When we consider 
the model of many fermion flavors, another current-current type four fermion 
contact term
$\sum_{a,b}\bar{\psi}_{a}\gamma^{\mu}\psi_{b}\,\bar{\psi}_{b}
\gamma_{\mu}\psi_{a}$ is allowed.} a well-known technique
to facilitate $1/N$-expansion is to rewrite Eq.(\ref{thi}) by introducing
an auxiliary vector field $A^{\mu}$ such as 
\begin{equation}\label{aux}
{\cal L}_{\rm aux}= \sum_{a}\bar{\psi}_{a}i\gamma^{\mu}(\partial_{\mu}
-\frac{i}{\sqrt{N}}A_{\mu})\psi_{a}+\frac{1}{2G}A_{\mu}A^{\mu}.
\end{equation}

Gauge-noninvariant as the above Lagrangian is, however one may be tempted
to regard $A_{\mu}$ as a gauge field. A systematic way to construct
the $U(1)$ gauge theory, but is gauge equivalent to Eq.(\ref{aux}) is
to elicit the fictitious Goldstone degree (or equivalently the St\"{u}ckelberg
field) based on the principle of hidden local symmetry 
\cite{BKY,KK,IKSY,Kon}:
\begin{equation}\label{hls}
{\cal L}_{\rm HLS}=\sum_{a}\bar{\psi}_{a}i\gamma^{\mu}D_{\mu}\psi_{a}+
\frac{1}{2G}(A_{\mu}-\sqrt{N}\partial_{\mu}\phi)^{2}.
\end{equation}
It is obvious that Eq.(\ref{hls}) possesses a $U(1)$ gauge symmetry and
the gauge-fixed (unitary gauge) form of it exactly coincides with
Eq.(\ref{aux}), so does the original Thirring model in Eq.(\ref{thi}).
The gauge theory formulation of Thirring model
in Eq.(\ref{hls}) looks like quantum
electrodynamics (QED) + scalar QED with
kinetic term of the gauge field and the Higgs degrees of freedom truncated
in the tree level.
Now that we find a gauge-invariant formulation of the Thirring model, we
have the privilege to choose the gauge appropriate for our particular
purpose. Here, instead of the unitary gauge notorious for loop
calculations, let us consider the nonlocal $R_{\xi}$ gauge at the Lagrangian
level
\begin{equation}
\label{GF}
{\cal L}_{\rm GF}=-\frac{1}{2}\Bigl(\partial_{\mu}A^{\mu}+\sqrt{N}
\frac{\xi(\partial^2)}{G}\phi\Bigr){1 \over \xi(\partial^2)}\Bigl(
\partial_{\nu}A^{\nu}+\sqrt{N}\frac{\xi(\partial^2)}{G}\phi\Bigr),
\end{equation}
where the gauge fixing parameter $\xi$ has the momentum- (derivative-)
dependence. It is straightforward to prove the possession of the BRS symmetry
in spite of the nonlocality of $\xi$ and thereby it guarantees the S-matrix
unitarity. Another intriguing point is that, in the combined Lagrangian of
Eqs. (\ref{hls}) and (\ref{GF}), the fictitious Nambu-Goldstone
boson $\phi$ is completely decoupled independently of the specific form of
$\xi(\partial^{2})$. Though we introduced $U(1)$ gauge symmetry by use of the
flavor-singlet current, the hidden local symmetry can easily be extended
to the non-Abelian one, {\it e.g.} $U(n)_{\rm global}\times U(n)_{\rm local}$
symmetry by use of the ``$U(n)/U(n)$'' nonlinear sigma model :
\be
\label{non}
{\cal L}_{Non-Ab.}=\sum_{a}\bar{\psi_a}i\gamma^{\mu}D_{\mu}
\psi_{a} -{N \over G}
 tr \left[ (D_{\mu} u\cdot u^{\dagger})^2 \right],
\ee
where
$A_{\mu}=A_{\mu}^{\alpha} T^{\alpha}$,
and $u=e^{i\phi}, \phi = \phi^{\alpha}T^{\alpha} $,
with $T^{\alpha}$ being the  $U(n)$ generators.
Actually, Eq.(\ref{non}) is gauge equivalent to
the Thirring model having the interaction
\be
-\frac{G}{2N}\sum_{a,b,\alpha}
(\bar{\Psi}_{a}\gamma^{\mu}T^\alpha \Psi_{a})\,
(\bar{\Psi}_{b}\gamma_{\mu}T^\alpha \Psi_{b}).
\ee
In contrast to the $U(1)$ case, however,
 the fictitious NG bosons
$\phi$ in the non-Abelian case are not
 decoupled even in the $R_{\xi}$ gauge,
which would make the analysis at quantum level rather complicated.

Now that we have reformulated
the Thirring model as a gauge theory,
we can further gain an insight
into the theory by using a technique
inherent to the gauge theory, namely,
the dual transformation \cite{Sav}.
We first consider the path integral
for the Lagrangian (\ref{hls}), and linearize the ``mass term" of gauge field by introducing an auxiliary field
$C_{\mu}$. Through
an integration over the scalar field $\phi$,\footnote{The scalar phase 
$\phi$ can
in fact be divided into two parts:
 $\phi=\Theta+\eta$,
 where $\Theta$ expressed by multi-valued function
 describes the topologically nontrivial sector, e.g.,
 the creation and annihilation of topological solitons,
 and $\eta$ given by single-valued function
 depicts the fluctuation around a given topological sector.
Inclusion of the topological sector $\Theta$ induces a topological
 interaction term \cite{Sav}, though we neglect $\Theta$
 contribution in this section, since we are interested in
 $\phi$ as the Nambu-Goldstone mode.} we obtain a delta functional for 
$\partial_{\mu} C^{\mu}$.
 If we pick up $C_{\mu}$ by use of
dual antisymmetric-tensor field
$H_{\mu_{1}\cdots\mu_{D-2}}$ of rank $D-2$,
 which satisfies the Bianchi identity,
we obtain the following path integral after integrating out the auxiliary field $C_{\mu}$:
\def\bpsi{\bar{\psi}}
\def\cdbm{D_\mu}
\bea
 Z_{\rm Dual} &=& \int[dH_{\mu_{1}\cdots\mu_{D-2}}]
                  [dA_{\mu}][d\bpsi_{a}][d\psi_{a}]
                \exp i\int d^{D}\!x\biggl\{
                      \sum_{a}\bpsi_{a}i\gamma^{\mu}\cdbm\psi_{a}
 \nonumber\\
          & &\hspace{0cm}
              +\frac{(-1)^{D}}{2(D-1)}
                     H_{\mu_{1}\cdots\mu_{D-1}} H^{\mu_{1}\cdots\mu_{D-1}}
              +\frac{1}{\sqrt{G}}\epsilon^{\mu_{1}\cdots\mu_{D}}A_{\mu_{1}}
               \partial_{\mu_{2}}H_{\mu_{3}\cdots\mu_{D}}
                                   \biggr\}
 \label{deltaf} \\
          &=& \int[dH_{\mu_{1}\cdots\mu_{D-2}}][d\bpsi_{a}][d\psi_{a}]\;
               \delta(\sum_{a} \frac{1}{\sqrt{N}}
                      \bpsi_{a} \gamma^{\mu_{1}} \psi_{a}
                     +\frac{1}{\sqrt{G}}\epsilon^{\mu_{1}\cdots\mu_{D}}
                      H_{\mu_{2}\cdots\mu_{D}})
 \nonumber\\
          & & \hspace{0cm}
               \exp i\int d^{D}\!x\biggl\{
                     \sum_{a} \bpsi_{a} i \gamma^{\mu} \partial_{\mu} \psi_{a}
                    +\frac{(-1)^{D}}{2(D-1)}
                     H_{\mu_{1}\cdots\mu_{D-1}}H^{\mu_{1}\cdots\mu_{D-1}}
                                  \biggr\},
 \label{eqn:dual}
\eea
where $H_{\mu_{1}\cdots\mu_{D-1}}
=\partial_{\mu_{1}}H_{\mu_{2}\cdots\mu_{D-1}}
-\partial_{\mu_{2}}H_{\mu_{1}\mu_{3}\cdots\mu_{D-1}}+\cdots+(-1)^{D}
\partial_{\mu_{D-1}}H_{\mu_{1}\cdots\mu_{D-2}}$.
The Lagrangian (\ref{eqn:dual}) describes $N$ ``free'' fermions and
a ``free'' antisymmetric tensor field of rank $D-2$
 which are, however,
 constrained through the delta functional.
 This implies that the dual field is actually
 a composite of the fermions.

Let us write down the discretized Lagrange density for the Thirring model
on the 3D Euclidean lattice. Of course, it looks not so good way to use
the original Lagrange density with four-fermion contact term in Eq.(\ref{thi})
directly.
Therefore, one way is to discretize Eq.(\ref{aux}): \cite{DH} 
\begin{equation}
{\cal L}_{DH}=\sum_{a}\Phi^{\dagger}_{a}(\tilde{M}^{\dagger}\tilde{M})^{-1}
\Phi_{a}-\frac{N}{2}\beta\sum_{\mu}\tilde{\theta}^{2}_{\mu}(x),
\end{equation}
where \(\Phi\) is the pseudo fermion, \(\tilde{\theta}_{\mu}\) is the vector
auxiliary field, $\beta = 1/G\Lambda$ is the inverse of the dimensionless
coupling rescaled by the ultraviolet cutoff $\Lambda$ which is identified 
with the
inverse of the lattice size $a$, $1/\Lambda=a$, and 
\begin{equation}\label{tmxy}
M_{x,y}=\frac{1}{2}\sum_{\mu}\eta_{\mu}(x)[(1+i\tilde{\theta}_{\mu}(x))
\delta_{y,x+\mu} -(1-i\tilde{\theta}_{\mu}(x-\mu))\delta_{y,x-\mu}]
+m\delta_{x,y}.
\end{equation}
In Eq.(\ref{tmxy}), $\eta_\mu (x)$ is staggered fermion phase factor, and $m$ a finite
bare fermion mass term. The other method is to adopt the gauge theory formulation and take
advantage of various benefits coming from the gauge theory formulation: 
\cite{KiKi} 
\begin{equation}\label{latac}
{\cal L}_{L}=\sum_{a}{\Phi^{\dagger}}_{a}(M^{\dagger}M)^{-1}\Phi_{a}
-N\beta\sum_{\mu}\cos{(\phi(x+\mu)-\phi(x)+\theta_{\mu}(x))},
\end{equation}
where \(\Phi\) is the pseudo fermion and \(\theta_{\mu}\) is the gauge field,
\begin{equation}\label{mxy}
M_{x,y}=\frac{1}{2}\sum_{\mu}\eta_{\mu}(x)(e^{i\theta_{\mu}(x)}\delta_{y,x+\mu}
-e^{-i\theta_{\mu}(x-\mu)}\delta_{y,x-\mu})+m\delta_{x,y}.
\end{equation}
Here $\theta_{\mu}(x)$ is a gauge field which appears in the Lagrangian
in terms of the link variable $\exp(i\theta_{\mu})$.
Note that the second term in Eq.(\ref{latac}) is so-called the annealed XY model
with the gauge connection between the nearest neighbors in statistical 
mechanics.

\section{Bosonization of the Thirring Model}
\def\pbn{\partial_\nu}
\def\fsl#1{\setbox0=\hbox{$#1$}           
   \dimen0=\wd0                                 
   \setbox1=\hbox{/} \dimen1=\wd1               
   \ifdim\dimen0>\dimen1                        
      \rlap{\hbox to \dimen0{\hfil/\hfil}}      
      #1                                        
   \else                                        
      \rlap{\hbox to \dimen1{\hfil$#1$\hfil}}   
      /                                         
   \fi}
\newcommand{\dfrac}[2]{\frac{\strut \displaystyle{#1}}%
{\strut \displaystyle{#2}}}
\def\tint{\int d^2\! x\,\,}

\indent\indent In (1+1) dimensions the relation in the delta functional
in Eq.(\ref{eqn:dual})
 implies nothing but the bosonization condition
 in the scheme of path integral, i.e.,
 $
 \frac{1}{\sqrt{G}}\epsilon^{\mu\nu} \pbn H
 \approx \frac{-1}{\sqrt{N}} \bpsi_a \gamma^\mu \psi_a
 $.
Integrating the fermions in Eq.(\ref{deltaf}),
 we obtain an effective theory which consists of a pseudoscalar
 and a vector gauge fields:
\def\abm{A_\mu}
\def\ptm{\partial^\mu}
\def\pbm{\partial_\mu}
\be
 Z_{2D}=\int[dH][d\abm] \left( \det i\fsl{D} \right)^N \hspace{1mm}
         \exp i\tint\biggl\{
                           \frac{1}{2} \ptm H\pbm H
                          +\frac{1}{2 \sqrt{G}}
 H \epsilon_{\mu\nu} F^{\mu\nu}
                    \biggr\}.\label{eqn:2D}
\ee
The second term of the action in
Eq.(\ref{eqn:2D}) is the (1+1) dimensional
 analogue of axion term which
is the interaction term between the scalar
 and the gauge fields and takes the form
$H\epsilon_{\mu\nu\rho\sigma}F^{\mu\nu}F^{\rho\sigma}$ in (3+1)
dimensions.
Though the computation of fermionic determinant with
regularization generates the Abelian chiral anomaly,
this problem is resolved by the constant
shift of scalar field $H$ in axion term.
Since the fermionic determinant
is computed in an exact form, i.e.,
\def\atm{A^\mu}
\def\atn{A^\nu}
\def\ggbmn{g_{\mu\nu}}
 $$
 -i N \ln
    \frac{\det i\fsl{D}}{\det i\fsl{\partial}}
  = \frac{1}{2\pi}\tint\atm
 (\ggbmn-\frac{\pbm\pbn}{\partial^{2}}) \atn,
 $$
 the integration over $\abm$ gives a free massless scalar theory
 as the bosonized Thirring model
\be\label{boson}
 Z_{\rm boson}=\int[dH]\exp
i\tint\frac{1}{2}(1+\frac{\pi}{G})\pbm H\ptm H .
\ee
If $G$ is in the region $-\pi<G<0$,
 the energy per unit volume is unbounded below and hence
 the (1+1)-dimensional Thirring model with
 coupling constant $G$ ($G>0$ or $G<-\pi$)
is well-defined.

Though there have been several attempts to interpret the relation in 
Eq.(\ref{eqn:dual}) as that for the bosonization in arbitrary dimensions
\cite{FS}, their achievement is unclear yet since the fermionic determinant
in spacetime dimensions more than (1+1)D produces nonlocal terms of the dual
field.

\section{Dynamical Symmetry Breaking}
\indent\indent In (2+1) dimensions, our Lagrangian in Eq.(\ref{hls}) is 
invariant under the parity
\begin{equation}
\psi_{a}(x)\mapsto\psi_{a}^{'}(x^{'})=i\gamma^{3}\gamma^{1}\psi_{a}(x),
\;\;\;A_{\mu}(x)\mapsto A_{\mu}^{'}(x^{'})=(-1)^{\delta_{\mu 1}}A_{\mu}(x).
\end{equation}
and the so-called global ``chiral'' transformation
\begin{equation}
\psi_{a}\mapsto\psi^{'}_{a}=\Bigl(\exp\bigl(i\omega^{i\alpha}
\frac{\Sigma^{i}}{2}\otimes T^{\alpha}\bigr)\psi\Bigr)_{a},
\end{equation}
where $\Sigma^{0}=I,\;\Sigma^{1}=-i\gamma^{3},\;\Sigma^{2}=\gamma^{5},\;
\Sigma^{3}=-\gamma^{5}\gamma^{3}$ and $T^{\alpha}$ denote the generators
of $U(N)$.
The question we shall address from now on is ``which symmetry is broken
dynamically?'' In concern with the parity, first issue is whether one can
take the regularization to keep both the $U(1)$ gauge symmetry and
the parity or not. Since our gauge action in tree level has
the parity-conserving mass and the number of two component Dirac fermion
species is even, the parity need not be violated by appropriate regulator.
For example, the introduction of parity-conserving Pauli-Villars regulator
leads to the parity-invariant effective action for the gauge field as have done
in (2+1)D quantum electrodynamics(QED${}_{3}$) \cite{Red}. Another question
is whether
the pattern of dynamically-generated fermion mass involves the parity
violating mass ($-m\bar{\psi}_{a}\gamma^{5}\gamma^{3}\psi_{a}$) or not.
Though at this stage we do not yet know whether the dynamical symmetry
breaking really occurs or not, in this gauge-invariant formulation of the
Thirring model, such symmetry breaking pattern is proven to be energetically
unfavorable by using the exact argument in Ref. [10]. Namely, since
the tree-level gauge action corresponding to Eq.(\ref{hls}) is real and
positive semi-definite in Euclidean space, energetically favorable is
a parity conserving configuration consisting of half the 2-component fermions
acquiring equal positive masses and the other half equal negative masses.

According to the above arguments, the pattern of symmetry breaking we shall
consider is not the parity but the chiral symmetry, i.e., of which the
breaking is $U(2N)\rightarrow U(N)\times U(N)$. Thus we investigate
the dynamical mass of the type $m\bar\psi \psi$ in the Schwinger-Dyson
equation, giving
\begin{eqnarray}
\lefteqn{(A(-p^2)-1)\fsl{p}-B(-p^2)=}&&\nonumber\\
&&-\frac{1}{N}\int \frac{d^D q}{i(2 \pi)^D}
\gamma_{\mu}\dfrac{A(-q^2)\fsl{q}+B (-q^2)}{A^2(-q^2)q^{2}-B^{2}(-q^2)}
\Gamma_{\nu}(p,q)\;iD^{\mu \nu}(p-q),\label{sdeq}
\end{eqnarray}
where the full fermion propagator is written as $S(p)=i[A(-p^2)
\fsl{p}-B(-p^2)]^{-1}$, and $\Gamma_{\nu}(p,q)$ and $D_{\mu \nu}(p-q)$
denote the full vertex function and the full gauge boson propagator,
respectively. Task is, by employing some appropriate approximations, to
reduce Eq.(\ref{sdeq}) to the tractable integral equation for the mass
function $M(-p^2) = B(-p^2)/A(-p^2)$. First, we here adopt the $1/N$
expansion for $\Gamma_{\nu}(p,q)$ and $D_{\mu \nu}(p-q)$ under a nonlocal
$R_{\xi}$ gauge, i.e., they are the bare vertex and the one-loop vacuum
polarization of massless fermion loop at the $1/N$ leading order.
Then the Schwinger-Dyson equation (\ref{sdeq}) becomes the coupled integral
equations for
$A(-p^2)$ and $B(-p^2)$. They support a trivial solution $A(-p^2)=1$ and
$B(-p^2)=0$ at the $1/N$ leading order, however, as was realized in
QED${}_{3}$, we expect to find a nonperturbative nontrivial solution
by examining them for finite $N$.

A way is, by use of the freedom of gauge choice, to require $A(-p^{2})=1$
in a Schwinger-Dyson equation for $A(-p^{2})$. Then this gauge fulfills
the consistency
between the bare vertex approximation and the Ward-Takahashi identity
for the hidden local $U(1)$ symmetry (or the current conservation), i.e.,
$A(0)=1$. The specific form of the gauge is determined by a Schwinger-Dyson
integral equation for $A(-p^{2})$, and it reduces the coupled Schwinger-Dyson
equations into a single equation for $B(-p^{2})$ which turns out to be a
mass function, i.e.,
$M(-p^{2})=B(-p^{2})$:
\begin{equation}\label{sdeqb}
B(p^2)=\dfrac{1}{N}\int_0^{\Lambda^{D-2}}d(q^{D-2})K(p,q;G)
\dfrac{q^2 B(q^2)}{q^2+B^2(q^2)},
\end{equation}
where $\Lambda$ is ultraviolet cutoff and the kernel $K(p,q;G)$ is given by
\begin{equation}\label{ker}
K(p,q;G)={\textstyle \frac{1}{ (D-2) 2^{D-1} \pi^{(D+1)/2} \Gamma \left(
\frac{D-1}{2}\right)}}\int_0^\pi d\theta\sin^{D-2}\theta\;d(k^2)[D-\eta(k^2)],
\end{equation}
with $k^2 = p^2 + q^2 - 2 pq \cos \theta$.
The gauge fixing parameter $\xi$ is a function of $k^{2}$ such as
\begin{eqnarray}
\lefteqn{\eta(k^{2})=\frac{-\xi(k^{2})C_{D}^{\;-1}k^{D-2}+k^{2}}{
\xi(k^{2})G^{-1}+k^{2}}}&&\\
&=&(D-2)\left[\left(1+ \frac{Gk^{D-2}}{C_D} \right){}_{2}F_{1}
\left( \textstyle{ 1,1+\frac{D}{D-2},2+\frac{D}{D-2};-\frac{Gk^{D-2}}{C_D} }
\right)-1\right],
\end{eqnarray}
where $C_D^{-1}\equiv\frac{2\;tr\;I}{(4\pi)^{D/2}}\Gamma\left(2-
\textstyle{\frac{D}{2}}\right)B\left(\textstyle{\frac{D}{2}},\;
\textstyle{\frac{D}{2}}\right)$ and ${}_2 F_1(a,b,c;z)$ the hypergeometric
function.
Note that the kernel $K(p,q;G)$
is positive-definite for positive arguments $p$, $q$, $G$ and is symmetric
under the exchange of $p$ and $q$.

Now let us show the existence of the nontrivial solution for the
Schwinger-Dyson equation (\ref{sdeqb}) when $2<D<4$. Since we have in mind
the continuous phase transition, the solution of our interest is
the nontrivial solution which starts to exist without gap in the vicinity
of the phase transition point. Such a bifurcation point is identified by
the existence of an infinitesimal solution $\delta B(p^{2})$ around the trivial
solution $B(p^{2})=0$. In terms of dimensionless variables
($p=\Lambda x^{1/(D-2)},\;\delta B(p^{2})=\Lambda\Sigma(x),\;g=G/
\Lambda^{2-D}$), Eq.(\ref{sdeqb}) is reduced to a linearized integral equation
\begin{equation}\label{lineq}
\Sigma(x)=\frac{1}{N}\int_{\sigma_m}^{1}dyK(x^{1/(D-2)},y^{1/(D-2)};g)
\Sigma(y),
\end{equation}
where $\sigma_{m}=(m/\Lambda)^{D-2}\;(0<\sigma_{m}\leq 1)$ is the rescaled
infrared cutoff and in fact $m$ is nothing but the dynamically generated mass
by the normalization $m=\delta B(m^{2})$. We can rigorously prove that,
{\it if $N$ is equal to the maximal eigenvalue of the kernel} (\ref{ker}),
{\it then there exists a nontrivial
solution $\Sigma(x)$}. Hence, for a given $\sigma_{m}$, each line
$N(g,\sigma_{m})$ on $(N,g)$ plane depicts a line of equal
dynamically-generated mass $m=\Lambda\sigma^{1/(D-2)}_{m}$. Therefore, the
critical line is
defined by $N_{\rm cr}(g)=\lim_{\sigma_{m}\rightarrow 0}N(g,\sigma_{m})$
which separates the broken phase from the symmetric phase.
It is difficult to obtain the explicit form of the critical line $N_{\rm cr}(g)$
for arbitrary $g$, however we can get it in the limit of infinite four-fermion
coupling constant, $g\rightarrow\infty$. In this limit, the bifurcation
equation (\ref{lineq}) in (2+1)D is rewritten into a differential equation
\begin{equation}\label{difeq}
\frac{d}{dx}\Bigl(x^2 \frac{d\Sigma(x)}{dx}\Bigr)=-\frac{32}{3\pi^2 N}
\Sigma (x),
\end{equation}
plus the infrared boundary condition $\Sigma^{\prime} (\sigma_m)=0$ and
the ultraviolet one $\mathop{[}x\Sigma^{\prime}(x)+\Sigma(x)\mathop{]}
{}_{x=1}=0$.
When $N>N_{\rm cr}\equiv 128/3\pi^2$, there is no nontrivial solution of
Eq.(\ref{difeq}) satisfying the boundary conditions, while for $N<N_{\rm cr}$
the following bifurcation solutions exist:
\begin{equation}\label{eqn:sol}
\Sigma(x)=\frac{\sigma_m}{\sin(\frac{\omega}{2}\delta)}\Bigl(
\frac{x}{\sigma_m}\Bigr)^{-\frac{1}{2}}\sin\biggl\{\frac{\omega}{2}
\Bigl[\ln\frac{x}{\sigma_m}+\delta\Bigr]\biggl\},
\end{equation}
where $\omega\equiv\sqrt{N_{\rm cr}/N-1}$, $\delta\equiv 2\omega^{-1}\arctan
\omega$ and $\sigma_m$ is given by the ultraviolet boundary condition:
\begin{equation}\label{uvb}
\frac{\omega}{2}\Bigl(\ln\frac{1}{\sigma_m}+2\delta\Bigr)=n\pi,\;\;\;
\mathop{n = 1,2,}\cdots\;\;.
\end{equation}
The solution with $n=1$ is the nodeless (ground state) solution
whose scaling behavior is read from Eq.(\ref{uvb}):
\begin{equation}\label{scal}
\frac{m}{\Lambda}=e^{2 \delta}\exp\biggl[-\frac{2\pi}{\sqrt{N_{\rm cr}/N - 1}}
\biggr].
\end{equation}
The critical four-fermion number $N_{\rm cr}=128/3\pi^{2}$ is
the same as the one in QED${}_{3}$ with the nonlocal gauge.

It is turn to comment briefly on the dynamically generated mass of the gauge
boson and the dual transformation. The vector (gauge) boson is merely an
auxiliary field at the tree level, however it turns out to be propagating by
obtaining the kinetic term through fermion loop effect
when the fermion acquires the dynamical mass.
In (2+1) dimensions the pole mass $M_{V}$ of the dynamical gauge boson is given
by
\begin{equation}\label{vmass}
\frac{1}{2\pi}\biggl[\frac{4m^{2}+M_{V}^{2}}{2M_{V}}\tan^{-1}
\frac{M_{V}}{2m}-m\biggr]=G^{-1},
\end{equation}
which always satisfies a condition that $M_{V}<2m$. 
Furthermore, the dual gauge field $H_{\mu}$ in Eq.(\ref{deltaf}) shares 
exactly the same
pole structure with the gauge field $A_{\mu}$ irrespectively of the phase.

\vspace{5mm}

Y.K. would like to thank K. Yamawaki and J. Nishimura for their hospitality
during SCGT96.
This research was supported in part by the KOSEF(95-0702-04-01-3).

\end{document}